\documentclass[12pt]{iopart}
\usepackage{graphicx}

\begin{document}

\title[Heavy Ion Physics with ATLAS]{Heavy Ion Physics at the LHC with the ATLAS Detector}

\author{P Steinberg, on behalf of the ATLAS Collaboration}
\address{Brookhaven National Laboratory, Upton, NY 11973}
\ead{peter.steinberg@bnl.gov}












\begin{abstract}
The ATLAS detector at CERN will provide a high-resolution longitudinally-
segmented calorimeter and precision tracking for the upcoming study of heavy
ion collisions at the LHC ($\sqrt{s_{NN}}=5520$ GeV).  The calorimeter
covers $|\eta|<5$ with both electromagnetic and hadronic sections, while
the inner detector spectrometer covers $|\eta|<2.5$.  ATLAS will
study a full range of observables necessary to characterize the
hot and dense matter formed at the LHC.  Global measurements (particle
multiplicities, collective flow) will provide access into its
thermodynamic and hydrodynamic properties.  Measuring complete jets
out to 100's of GeV will allow detailed studies of energy loss and
its effect on jets.  Quarkonia will provide a handle on deconfinement
mechanisms.  ATLAS will also study the structure of the nucleon and
nucleus using forward physics probes and ultraperipheral collisions,
both enabled by segmented Zero Degree Calorimeters.

\end{abstract}


\section{Introduction: Heavy Ion Physics at the LHC}

Heavy ion physics at the LHC is the next natural step in the
evolution of the understanding of QCD.
This can be seen most clearly when one considers the dynamical
evolution of a heavy ion collision.
The study of particle multiplicities and monojets give insight
into high density QCD and the parton structure of the nucleus
via models based on parton saturation,
such as the color glass condensate\cite{Iancu:2003xm}.
Hard processes probe the very earliest phase of the collision
process via the production of jets, photons, and heavy quark
states from parton-parton interactions~\cite{Accardi:2004gp}.
Through the use
of these calibrated probes, one can study their modification
in nuclear collisions and thus learn about QCD in medium
as well as the medium itself~\cite{Adler:2006bv}.
By the study of particle yields in $\eta$, $\phi$ and $p_T$,
both inclusive and identified, one can make connections to
hydrodynamics (both ideal and not) and probe the equation
of state which encodes the relevant microscopic degrees of freedom
\cite{Huovinen:2006jp}.
Finally, the study of integrated yields and the comparison of
different hadron species and their decays gives a handle on
the thermal and statistical properties of the system
\cite{Braun-Munzinger:2001as}.

The suite of collision systems (p+p, A+A, p+A) 
and detectors (ALICE, ATLAS, and CMS) at the LHC are ideally
suited to address all of the above theoretical questions via precise
experimental measurements using phenomenological tools 
developed in the context of RHIC 
collisions~\cite{Accardi:2004gp,Wiedemann:2006yu}.
Here we discuss
the progress of the ATLAS heavy ion effort to ready itself
for Pb+Pb running in late 2008 or early 2009.
Previous progress in preparations for ATLAS running have been
reported in Refs.~\cite{Rosselet:2005sy,White:2005au,Takai:2004nm,Takai:2004ka,
Nevski:2004zn} and in the Letter of Intent~\cite{LoI}.
\footnote{
{\bf Note on Figures:}
Unless otherwise noted, the figures shown here were based on
studies using modified versions of ATLAS production software.
Thus, they should be considered ``ATLAS preliminary''.  
For completeness, we list the version of Athena software
used for each figure:  
Figure 2(left) used 12.0.31 with a new tracking algorithm while
Figure 2(right) and Figure 3 used 12.0.3.
Figures 4-6 used 11.0.41 and special jet algorithms.  
Figure 7(right) was produced with an unmodified 11.0.3.
}

\section{ATLAS Detector}

The ATLAS detector
is a powerful tool for studying the high multiplicity
of particles that emerge from the collision of protons and 
nuclei~\cite{:1999fq,unknown:1999fr}.
A particular strength of the ATLAS detector is the hermetic
liquid argon (LAr) electromagnetic (EM) calorimeter, which provides
excellent energy and position information on electrons and photons
via its longitudinally-segmented towers~\cite{unknown:1996fq}.  
Of course, there is also a sophisticated
inner tracking system for reconstructing charged tracks and a
large volume muon tracking system.  

\begin{figure}[t]
\begin{center}
\includegraphics[width=120mm]{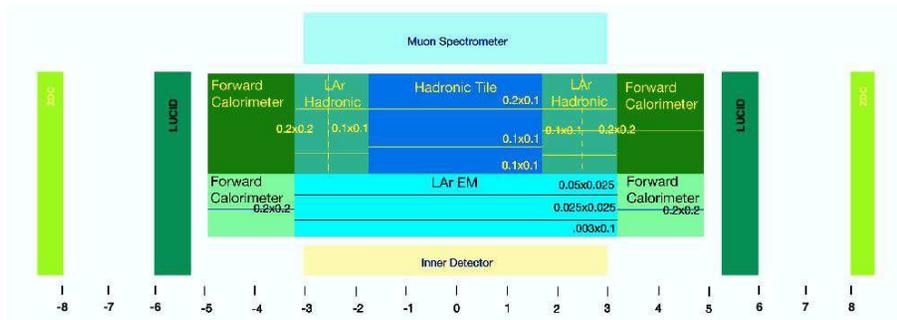}
\caption{
\label{fig:ATLASacc}
ATLAS acceptance in pseudorapidity.  All detectors have complete azimuthal
coverage.
}
\end{center}
\end{figure}

Like many modern collider detectors, ATLAS has hermetic azimuthal
coverage over a wide range in pseudorapidity.  The inner tracking
system covers $|\eta|<2.5$ with silicon pixels, silicon strips (SCT),
and a straw-tube transition-radiation tracker (TRT).
The electromagnetic calorimeter covers $|\eta|<3$ with
angular resolution depending on the layer ($\Delta\eta\times\Delta\phi = 0.003\times 0.1, 0.025 \times 0.025, 0.05 \times 0.025$).
The barrel hadronic tile calorimeter covers ($|\eta|<1.8$) with 
towers of $0.1 \times 0.1$.  In the forward direction, the hadronic
forward calorimeter has cells up to $0.2 \times 0.2$, and
a forward LAr calorimeter covers up to $\eta = 5$ with cells of
$0.2 \times 0.2$.

Additional detectors will extend the ATLAS acceptance farther into
the forward region.  The LUCID gas Cerenkov detector~\cite{Pinfold:2005sq} 
detects primary 
charged particle from $5.3<|\eta|<6$.  While it will be primarily 
purposed for luminosity monitoring, it should also provide a measure
of forward particle yields with a readout system that can resolve
multiple particles per tube.  More importantly for heavy ion physics
is the Zero Degree Calorimeter (ZDC) being designed and built especially
for ATLAS by a consortium of US groups~\cite{ATLASZDC}.  
This detector will be able
to detect forward neutron spectator fragments, which serves as
both a high-purity event trigger as well as an estimator for
the ``centrality'' of the collision (related directly to the
impact parameter).
However, its highly-segmented front EM section will also be able to 
measure the angle of neutral clusters.  This allows the reconstruction 
of neutral decays like $\pi^0$ and $\eta$, which will be discussed
below.

\section{Global Dynamics}

\begin{figure}[t]
\begin{center}
\includegraphics[height=60mm]{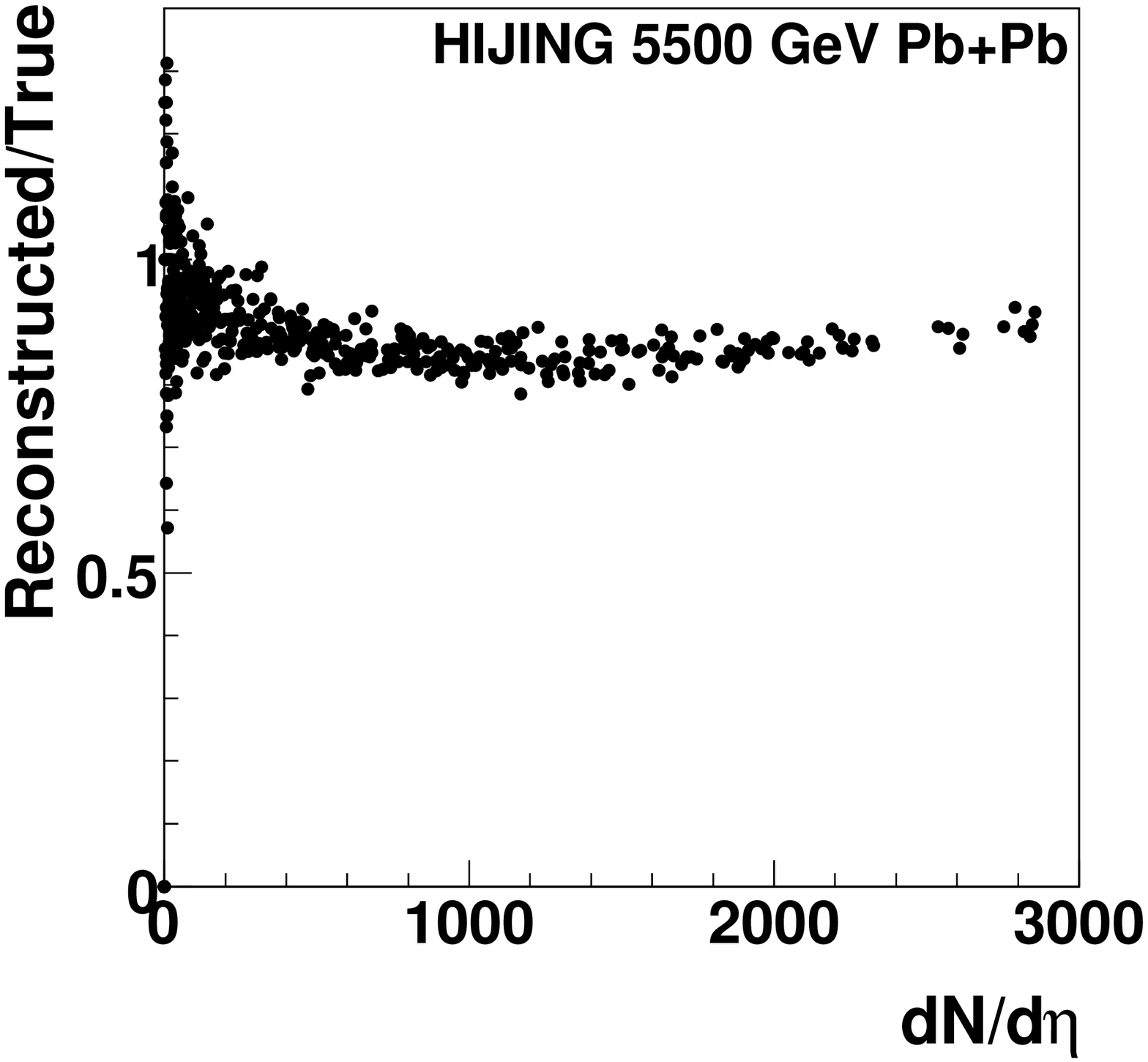}
\includegraphics[height=60mm]{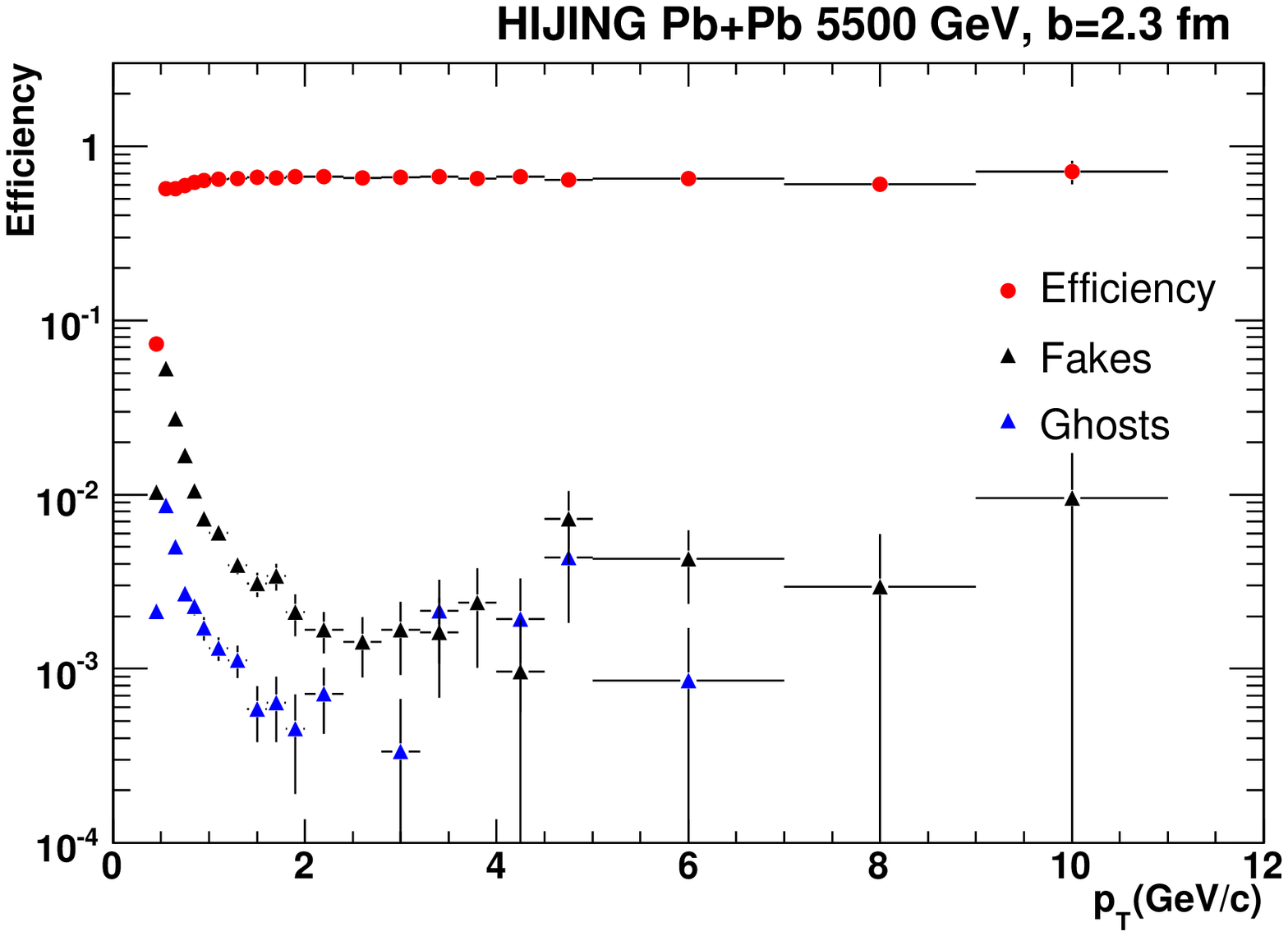}
\caption{
\label{fig:tracklets}
(left) Reconstruction of tracklets in the ATLAS pixel detectors.
(right) Tracking reconstruction efficiency, ghost rate, and fake rate
in central HIJING events.
}
\end{center}
\end{figure}

The most pressing issue for the early days at the LHC is to
establish the global features of heavy ion collisions.
This involves the estimation of the inclusive charged-particle
yield, both integrated and as a function of pseudorapidity,
to get a handle on the initial-state entropy production
which controls the hydrodynamic evolution as well as jet 
quenching~\cite{Baier:1996sk}.
It also entails studying the elliptic flow for
inclusive particles as a function of centrality, $p_T$ and 
pseudorapidity~\cite{Kolb:2003dz}.
One of the major questions for the LHC heavy ion program is
whether the presumed ``hydro limit'' has really been reached
in RHIC collisions, or if the magnitude of elliptic flow scaled
by the eccentricity ($v_2/\epsilon$)
will continue to increase with particle yield.

Estimating the particle density can be done in a variety of ways.
Even before the tracking system is fully commissioned, a reasonably-aligned 
pixel detector can be used to measure particle yields by the
``tracklet'' technique pioneered at RHIC by the PHOBOS 
experiment~\cite{Back:2000gw}.
This technique involves matching the angles of two pixel space points
with the estimated event vertex measured using the rest of the
inner detector.  While it is not as robust as the full tracking
procedure, it has a lower intrinsic $p_T$ cutoff and thus offers
quick access to the full yield.  Initial studies of tracklets
with full simulations,
shown in the left panel of Fig.~\ref{fig:tracklets},
show a good efficiency which is constant 
over a wide range of multiplicities emitted in $|\eta|<1$.
Of course the full inner detector will be indispensible for 
estimating particle yields and spectra.
The performance of early studies of the full tracking algorithm in
a heavy ion environment is shown in the
right panel of Fig.~\ref{fig:tracklets}. 
For central ($b=2.3$ fm) HIJING events, we find an
approximately constant efficiency of $\sim 70\%$
over a broad range in $p_T$ and low fake and ghost rates.

Estimating elliptic flow involves the calculation of the Fourier 
components of the measured angular distribution relative to the
``reaction plane'' (the angle $\Psi_{RP}$ seen by the neutron spectators)
or ``event plane'' (the angle $\Psi_{EP}$ seen by the emitted particles
themselves)~\cite{Ackermann:2000tr}.  The parameter $v_2$ is
defined via the expansion $dN/d\phi \propto 1 + 2 v_2 \cos[ 2(\phi-\Psi) ]$
where $\Psi = \Psi_R$ or $\Psi_E$, and is estimated by
calculating $v_2 = \langle \cos[ 2(\phi-\Psi) ]\rangle$ from experimental data.
\begin{figure}[t]
\begin{center}
\includegraphics[width=140mm]{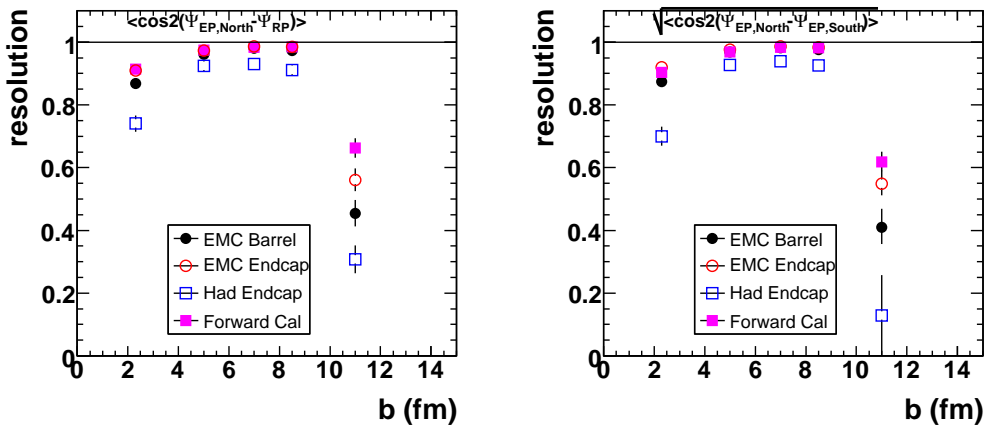}
\caption{
\label{fig:reso}
Two calculations of reaction plane resolution in ATLAS, relative
to the true reaction plane (left) and between symmetric subevents (right).
See text for definitions of these quantities.
Results are shown for different calorimeter subdetectors, which
cover different pseudorapidity regions (see Fig.~\ref{fig:ATLASacc}).
}
\end{center}
\end{figure}

The key figure of merit which controls the quality of the measurement
is the reaction plane resolution which is used to correct the experimental
data.
This can be estimated in an ideal case by the expression 
$\langle \cos (\Psi_{EP} - \Psi_{RP} ) \rangle$.
In a real measurement, it is typically assumed that the full event
sees the same reaction plane and the formula
$\sqrt{ \langle \cos[ 2(\Psi_{EP,N} - \Psi_{EP,S}) ] \rangle }$ is used
to estimate the reaction plane resolution by the measurement of
two symmetric subevents, one in the forward hemiphere and the
other in the backward hemisphere. 
Fig.~\ref{fig:reso} shows the ideal and subevent reaction plane
resolution as a function of impact parameter and subdetector,
each of which cover a different pseudorapidity region.
The resolutions are typically near 1 for most of the inelastic
cross section.  Even for the most challenging environments, e.g.
peripheral collisions with low multiplicity and central collisions
with a low $v_2$ signal, they are always above 0.3.
These high resolutions provide a major advance over RHIC measurements 
which had reaction plane resolutions typically {\it below} 
0.5~\cite{Adler:2002pu,Back:2002gz}.

\section{Jets}

The detailed study of fully-reconstructed jets will be
the major contribution of the LHC to the understanding of the
strongly-interacting QGP thought to be formed in heavy ion collisions
at RHIC.
Both the copious production of jets and the experimental acceptance
are unprecedented in the study of relativistic
heavy ions~\cite{Accardi:2004gp}.
The rates of hard processes in $p+p$
and $A+A$ collisions will increase dramatically relative to
RHIC energies.  Of course, so will the yields of uncorrelated soft
particles, which may well suffer enormous fluctuations if
copious minijet production indeed dominates the bulk 
particle production.
Thus, it is essential to have a calorimeter with a large acceptance in
$\eta$ and $\phi$ with excellent energy and position resolution,
to contain full jet, dijet, $\gamma$-jet and $Z$-jet events.  
Combining information from the different measurements 
will provide a good handle on the jet $E_T$ scale (e.g. from $\gamma$-jet
and $Z$-jet event) and 
fragmentation properties.

\begin{figure}[t]
\begin{center}
\includegraphics[width=70mm]{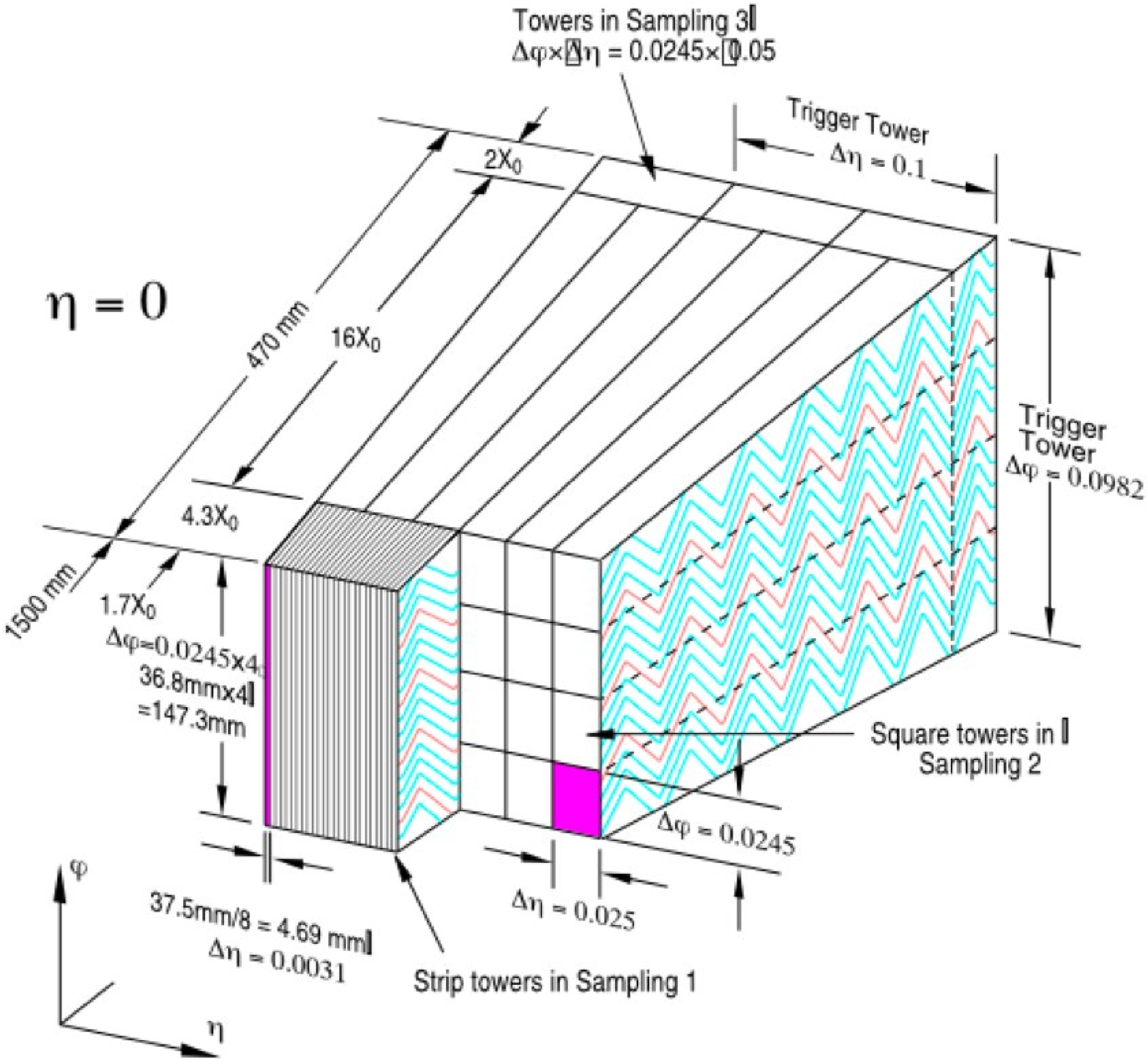}
\includegraphics[width=70mm]{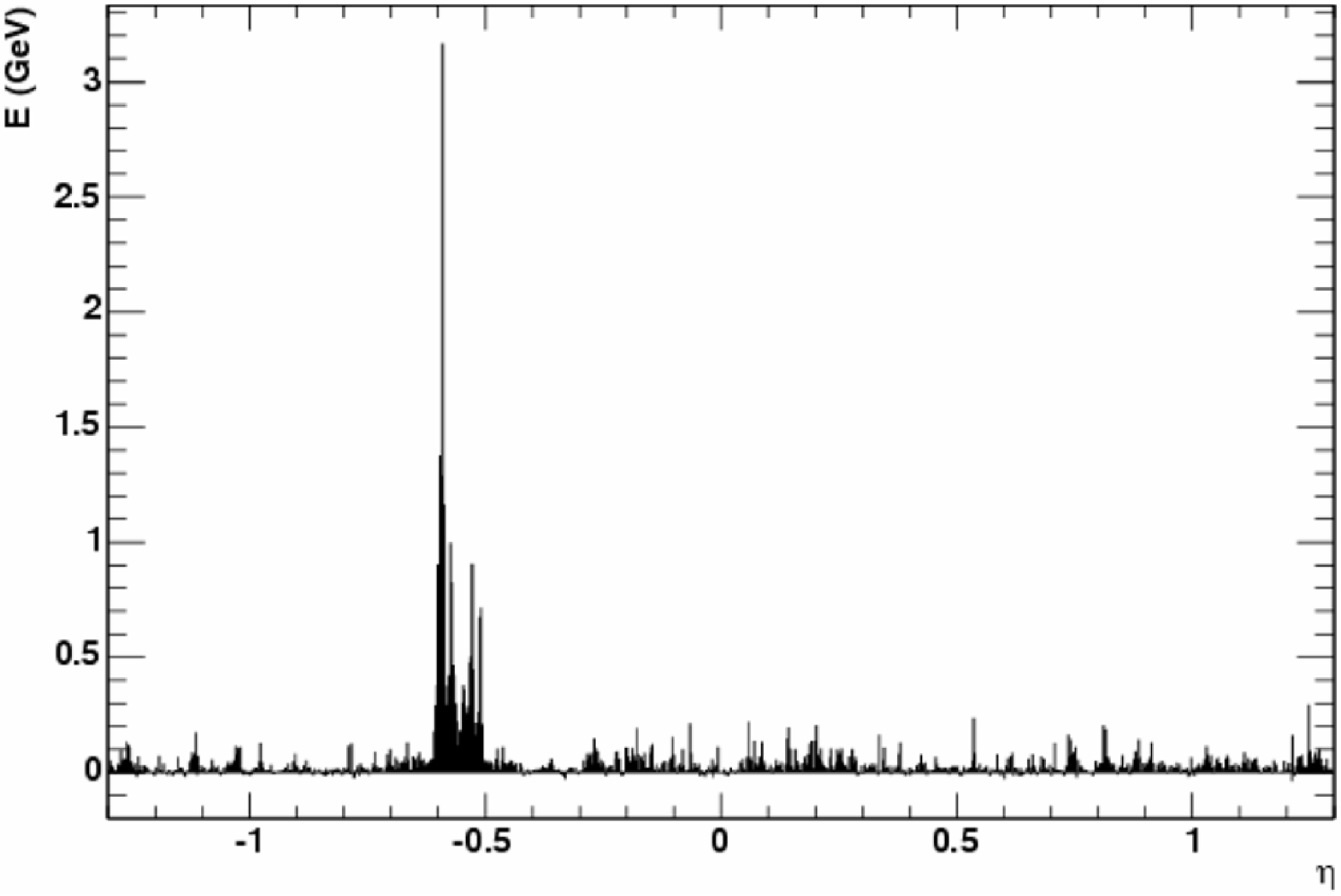}
\caption{
\label{fig:photon}
(left) Schematic of the ATLAS EMCAL longitudinal segmentation.
(right) Signals in the first EMCAL layer strips, showing a photon 
amidst uncorrelated HIJING background.
}
\end{center}
\end{figure}
As mentioned above, the longitudinal segmentation of the
ATLAS electromagnetic (EM) calorimeter, illustrated in
Fig.\ref{fig:photon} is unique at the
LHC and will be essential for jet physics at the LHC~\cite{unknown:1996fq}.
Detailed simulations have shown that 60\% of the
total energy (including both charged and hadronic energy)
ranges out in the first EMCAL layer.
Most of the hadronic component is charged and only leaves
MIPS in the first layer.  This means that photons,
especially those of several GeV and above, are easily
observed above the large central HIJING background in the first layer, as shown
in Fig.~\ref{fig:photon}.  This will dramatically enhance
ATLAS's ability to make direct photon measurements
by being able to reject even close decay photon pairs~\cite{ATL-PHYS-PUB-2005-018}.

\begin{figure}[t]
\begin{center}
\includegraphics[height=50mm]{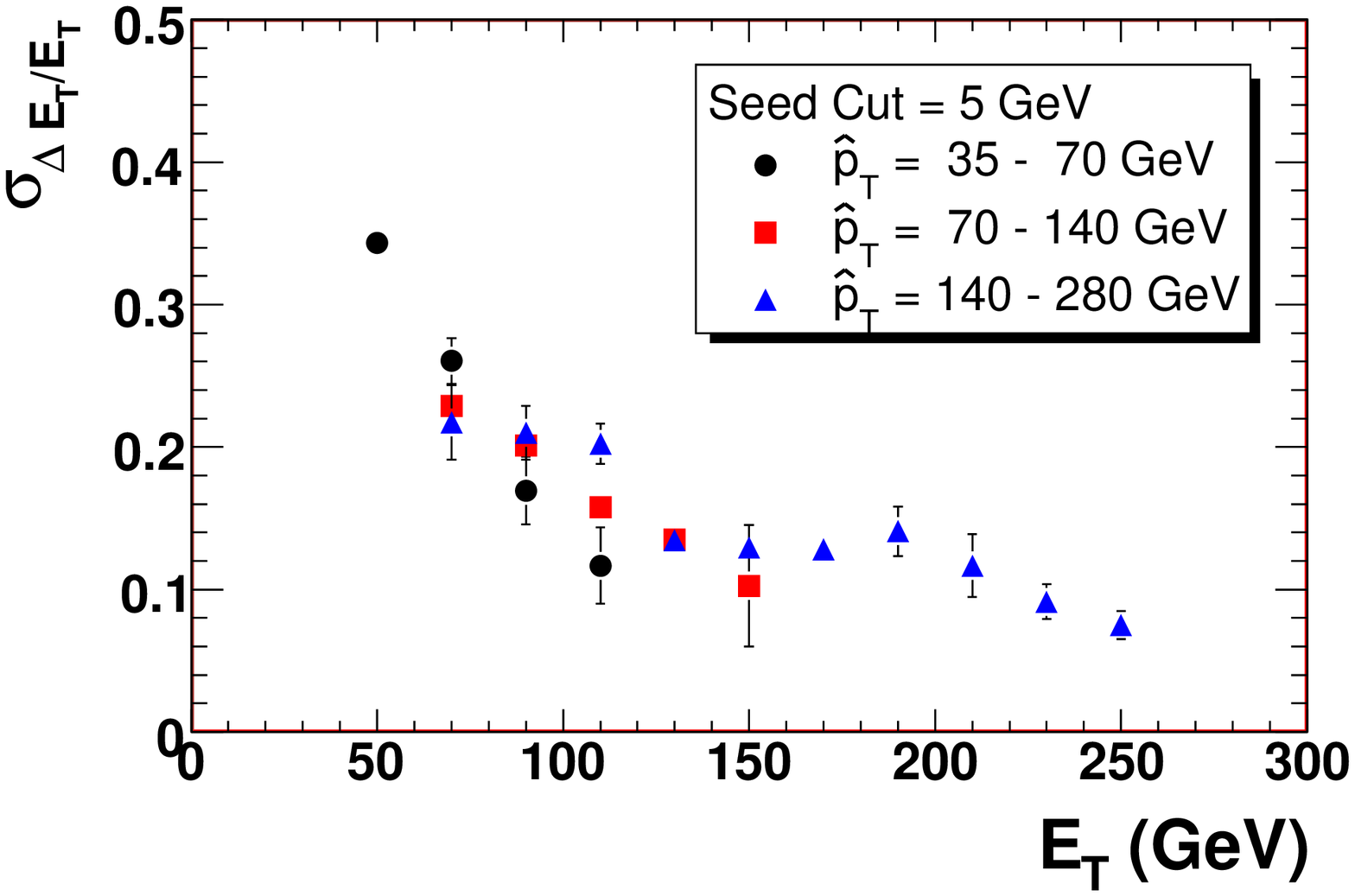}
\includegraphics[height=50mm]{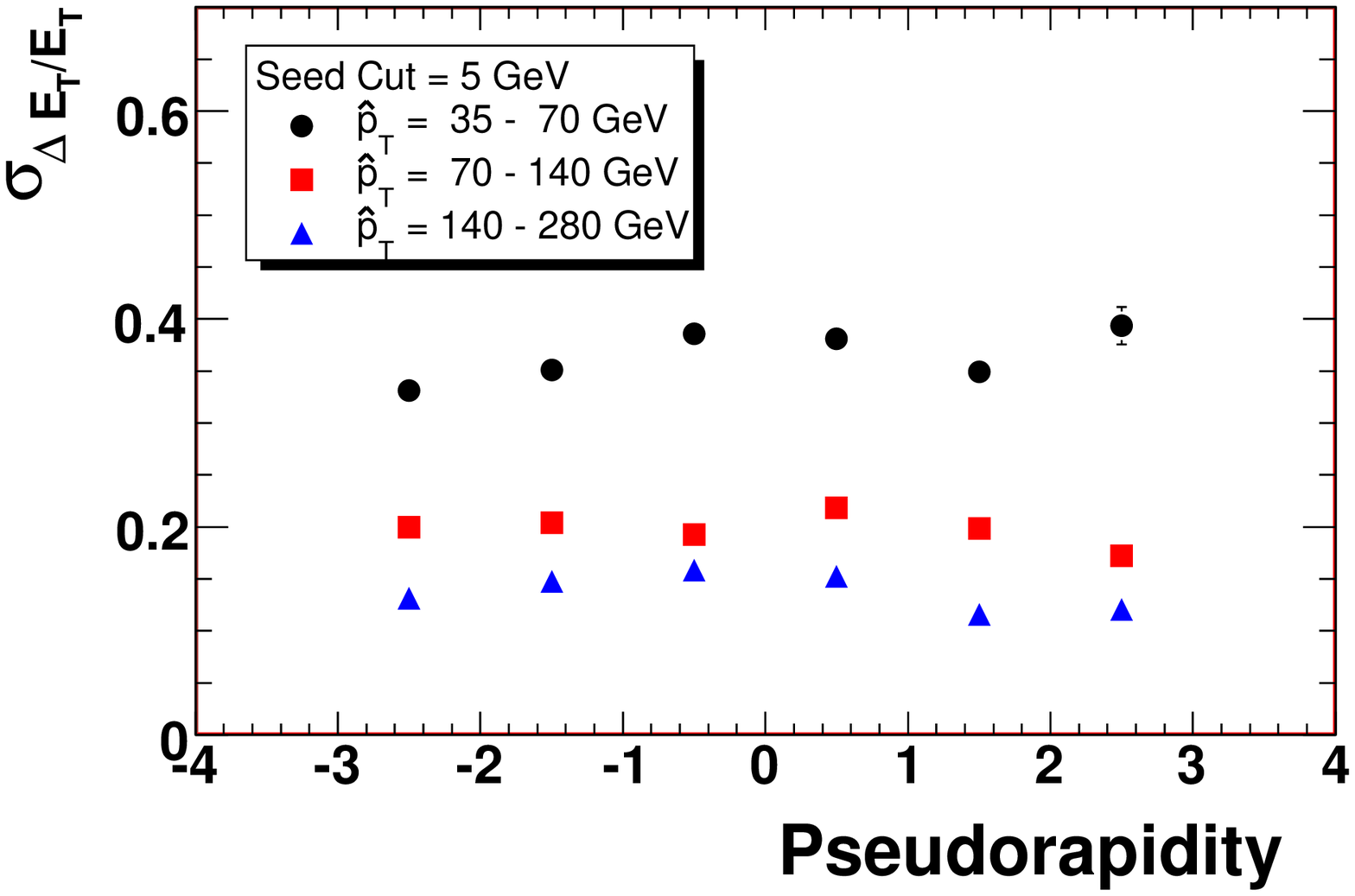}
\caption{
\label{fig:EnergyRes}
Energy resolution for jets in the ATLAS acceptance, as a function 
of jet $E_T$ and $\eta$.
}
\end{center}
\end{figure}

In order to study the ability of ATLAS to identify and reconstruct
jets, e.g. generated by PYTHIA, they are embedded into heavy ion events.
From these samples, various reconstruction procedures can be tested
and evaluated.
One procedure involves estimating the background by excluding
jet candidates and averaging the energy in the calorimeters within
a chosen tower size.  This average energy is subtracted and then
the standard ATLAS jet reconstruction algorithms are run on the
modified towers.  This allows heavy ion data analysis to take advantage
of progress with jet reconstruction algorithms and calibration.
The current results, applying the subtraction technique and the standard
ATLAS jet reconstruction algorithms down to 50 GeV,
are shown in 
Fig.\ref{fig:EnergyRes} for jets ranging from $E_T = 50-300$ GeV
and resolutions range from $\sim 25\%$ at the lowest energies
considered to $10-12\%$ at higher energies.  The resolution is
currently independent of $\eta$, as seen in Fig.~\ref{fig:EnergyRes}
for $|\eta|<2.5$. 

\begin{figure}[t]
\begin{center}
\includegraphics[width=70mm]{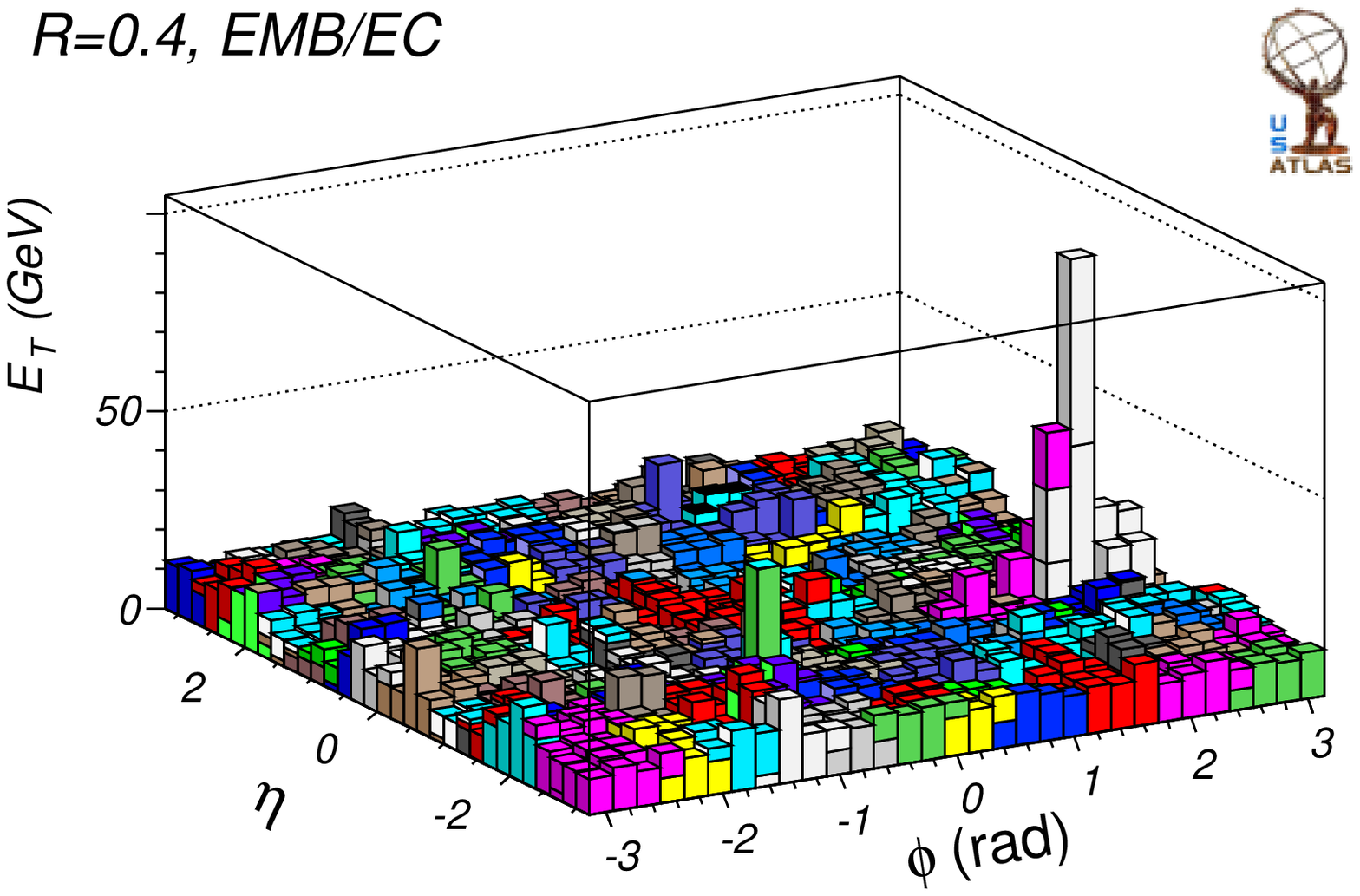}
\includegraphics[width=70mm]{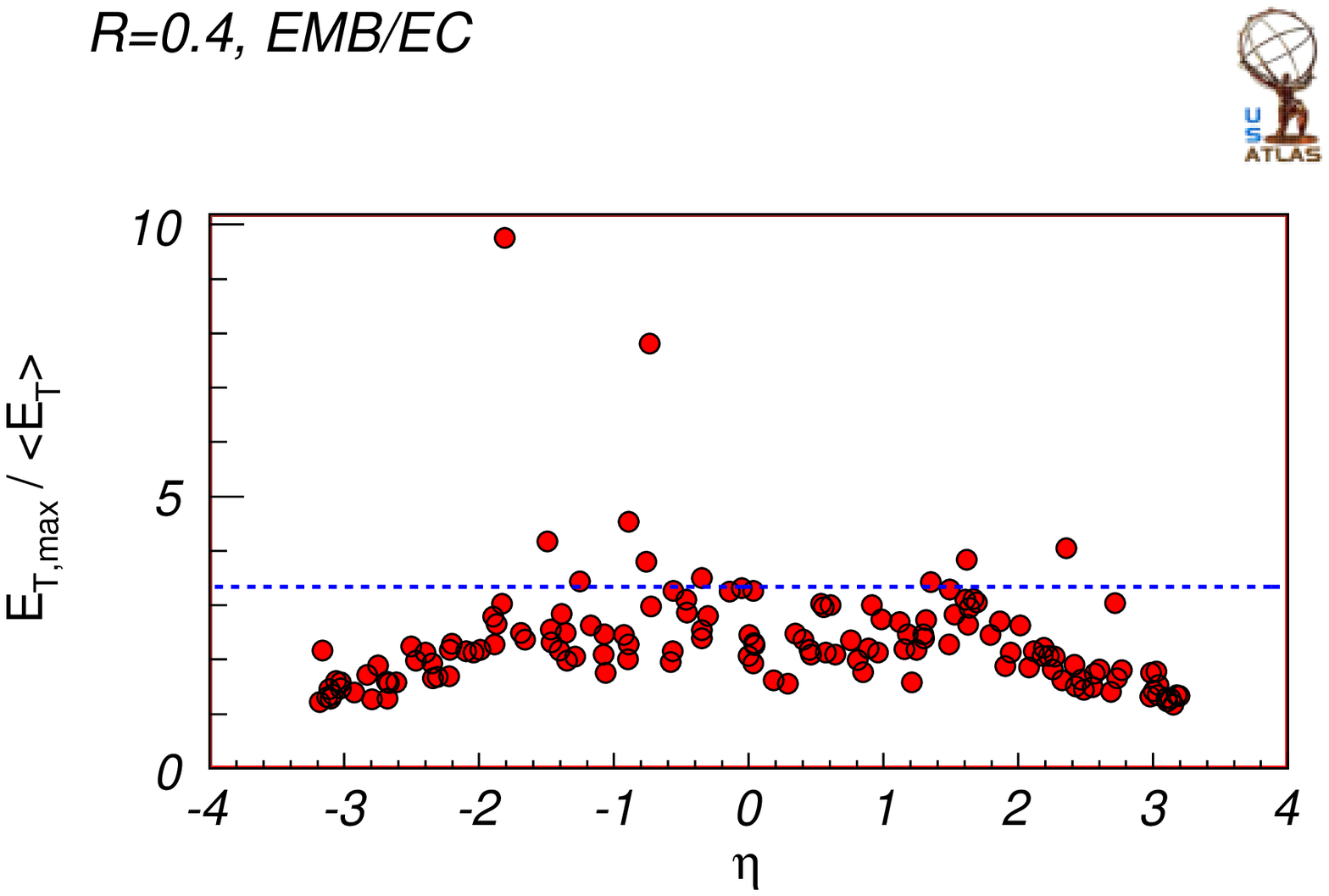}
\caption{
\label{fig:ktjets}
Extraction of jets from ATLAS calorimeters using the ``Fast $k_T$''
algorithm.
}
\end{center}
\end{figure}

Another technique currently under intense development is one applying
the ``Fast $k_T$'' algorithm~\cite{Cacciari:2005hq}
directly to heavy ion data without
a separate subtraction step.  In general, $k_T$ algorithms reconstruct
jets backwards along the fragmentation chain by combining particles
that minimize $d_{ij} = \mathrm{min}(k_{iT},k_{jT})R^2$
(where $R \equiv \sqrt{ \Delta \eta^2 + \Delta \phi^2}$) which
essentially encodes the $1/k^2_T$ probabilities of parton splitting.
While these algorithms are typically $O(N^3)$ (where $N$ is the
number of tracks or calorimeter clusters), Cacciari and Salam
have used the technique of Voronoi diagrams to reduce the problem
to $O(N \log N)$.  This allows the algorithm to be run quickly
even in central heavy ion events.  Early results are shown in 
Fig.~\ref{fig:ktjets}.  One can set a maximum radius for particles
to be clustered, e.g. $R_{max} = 0.4$
which groups the towers into many (e.g. 10's) of jet candidates.
These candidates can be characterized by various properties, 
e.g. maximum tower energy and average cell energy, as illustrated
for a single event in the left panel of Fig.~\ref{fig:ktjets}.
The ratio of these quantities for a single event's jet candidates 
is shown as a function of pseudorapidity in
Fig.~\ref{fig:ktjets}, where it is observed that jets are easily
distinguished from the rest of the background on an event-by-event
basis, and without a separate background-subtraction step.

\section{Quarkonia}

\begin{figure}[t]
\begin{center}
\includegraphics[width=70mm]{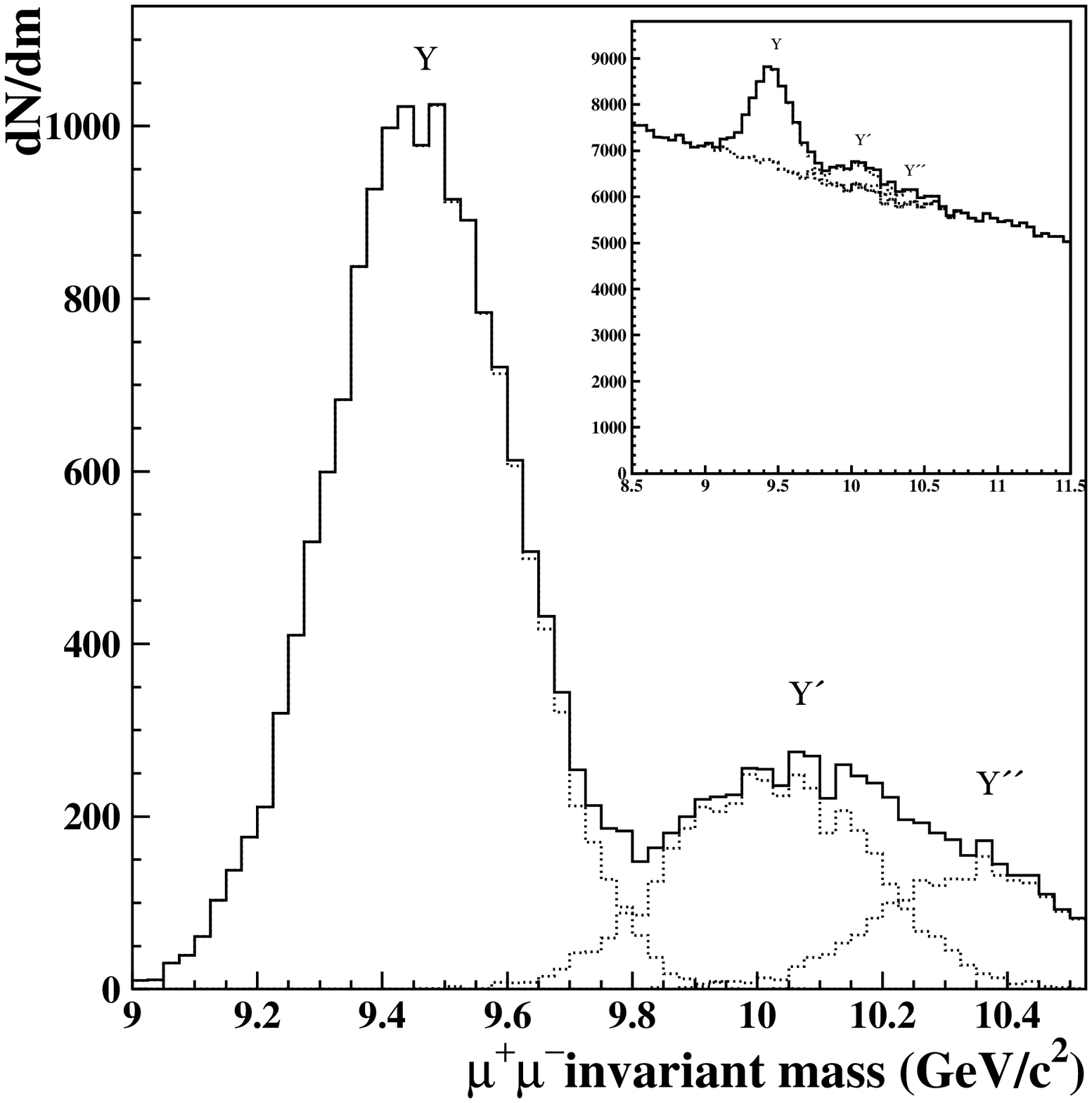}
\includegraphics[width=70mm]{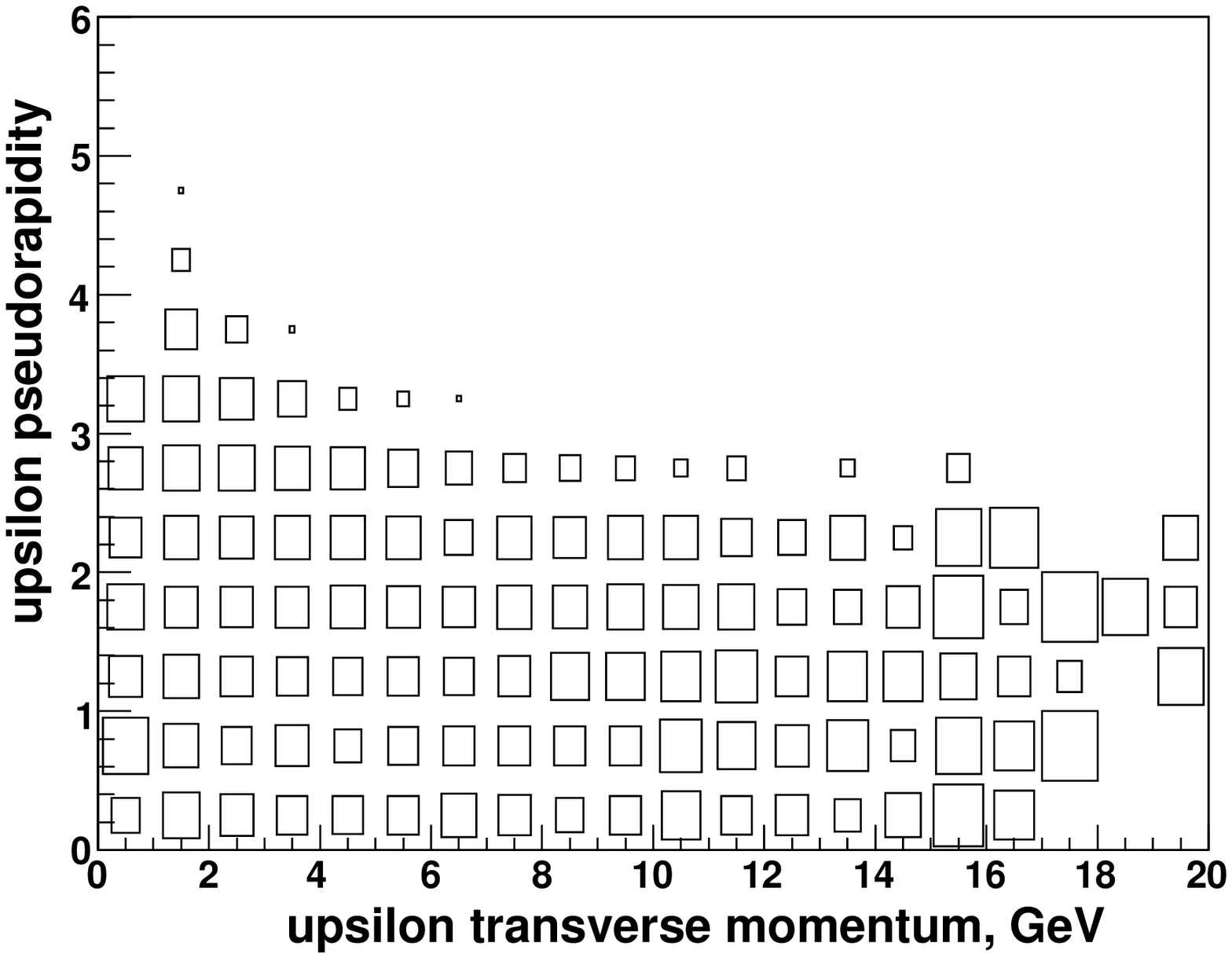}
\caption{
\label{fig:staco}
(left) Reconstruction of upsilons within the ATLAS muon spectrometer.
(right) Acceptance $\times$ efficiency for upsilons as a function
of pseudorapidity ($\eta$) and transverse momentum ($p_T$).
}
\end{center}
\end{figure}

The suppression of $J/\Psi$'s produced in heavy ion collisions
has become a major question arising from recent RHIC data.
When cast as $R_{AA}$, it is found that the suppression slightly
forward of mid-rapidity is quite similar in data from the
SPS ($\sqrt{s_{NN}}=17.3$ GeV, and RHIC ($\sqrt{s_{NN}}=200$ 
GeV)~\cite{Lajoie:2006}.
These two energies are an order of magnitude different, with
particle densities different by a factor of two, making the
similarity in the data quite puzzling.  The LHC will increase
$\sqrt{s_{NN}}$ by another factor of 27, which should shed some
light on the situation regardless whether the suppression 
patterns remain energy independent or undergo a dramatic change.

The ATLAS muon spectrometer is a high precision tracker covering
$|\eta|<2.5$ with full azimuth.  While it has unprecedented rapidity coverage
in heavy ion collisions, its design is optimized for very high energy
muons.  Thus, while it can measure muons of order 1 TeV, the material
budget of the inner detector and calorimeters make it difficult to
reconstruct muons below $p_T=3$ GeV/c.  This imposes minimum $p_T$ cuts on
$J/\Psi$ and $\Upsilon$ reconstruction since each muon needs to have
several GeV.

The left panel of Fig.~\ref{fig:staco} shows an ATLAS reconstruction
of upsilon mesons reconstructed in a high statistics sample of
$p+p$ collisions.  The mass resolution at present is 
$\sigma_{M_\Upsilon}\sim 120 MeV/c^2$ for $|\eta|<1$,
which is only slightly affected by
the presence of an uncorrelated heavy ion background.  The
right panel of Fig.~\ref{fig:staco} shows the ATLAS acceptance for
Upsilons as a function of pseudorapidity and transverse momentum.  
This figure shows a broad $\Upsilon$ acceptance that goes beyond the
nominal spectrometer resolution, stemming from the dimuon decay
kinematics, and out to very high $p_T$.  ATLAS will be
sensitive to quarkonia states over a wide kinematic range,
and will thus probe various aspects of deconfinement dynamics.

\section{Low $x$ Physics}

\begin{figure}[t]
\begin{center}
\includegraphics[width=70mm]{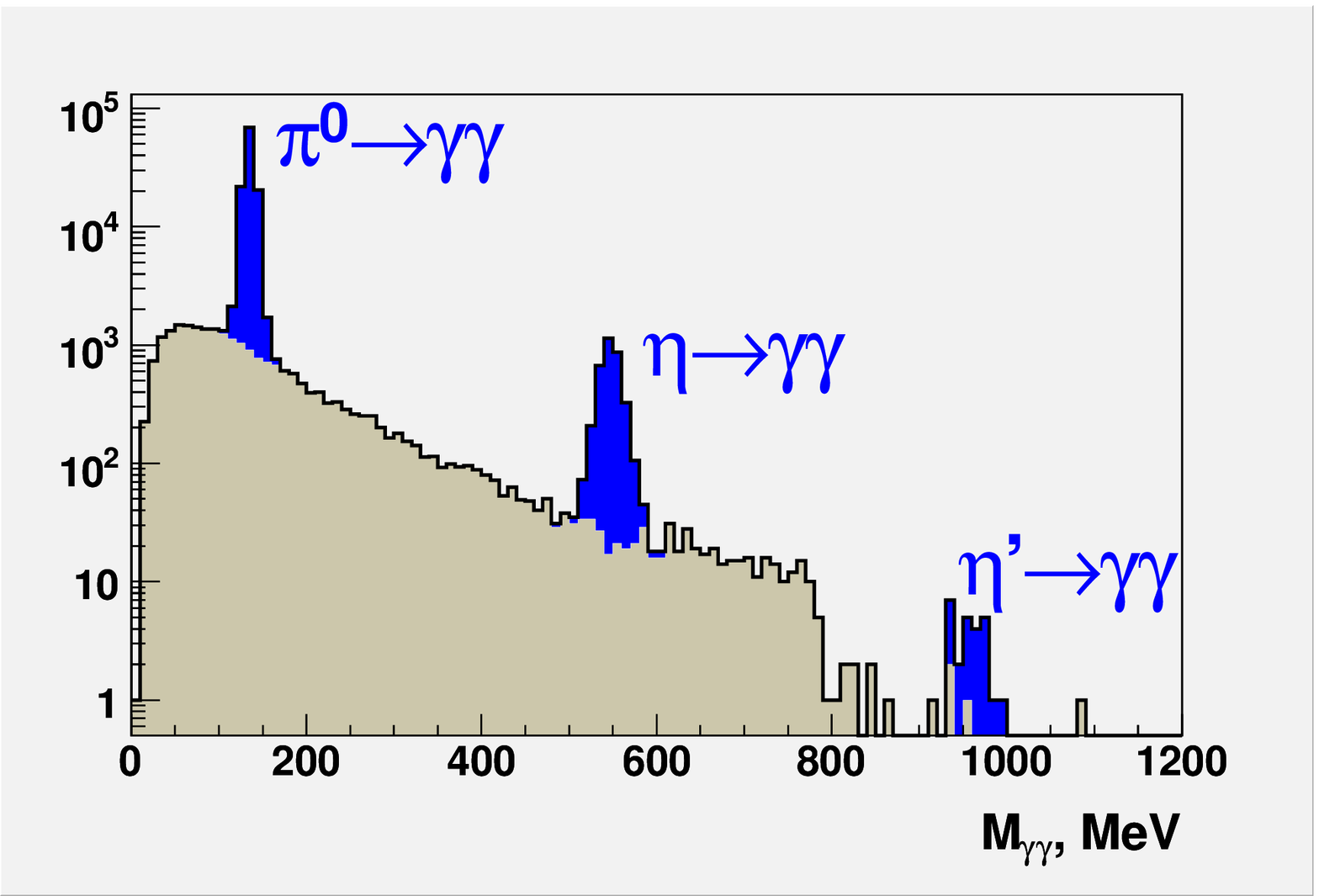}
\includegraphics[width=70mm]{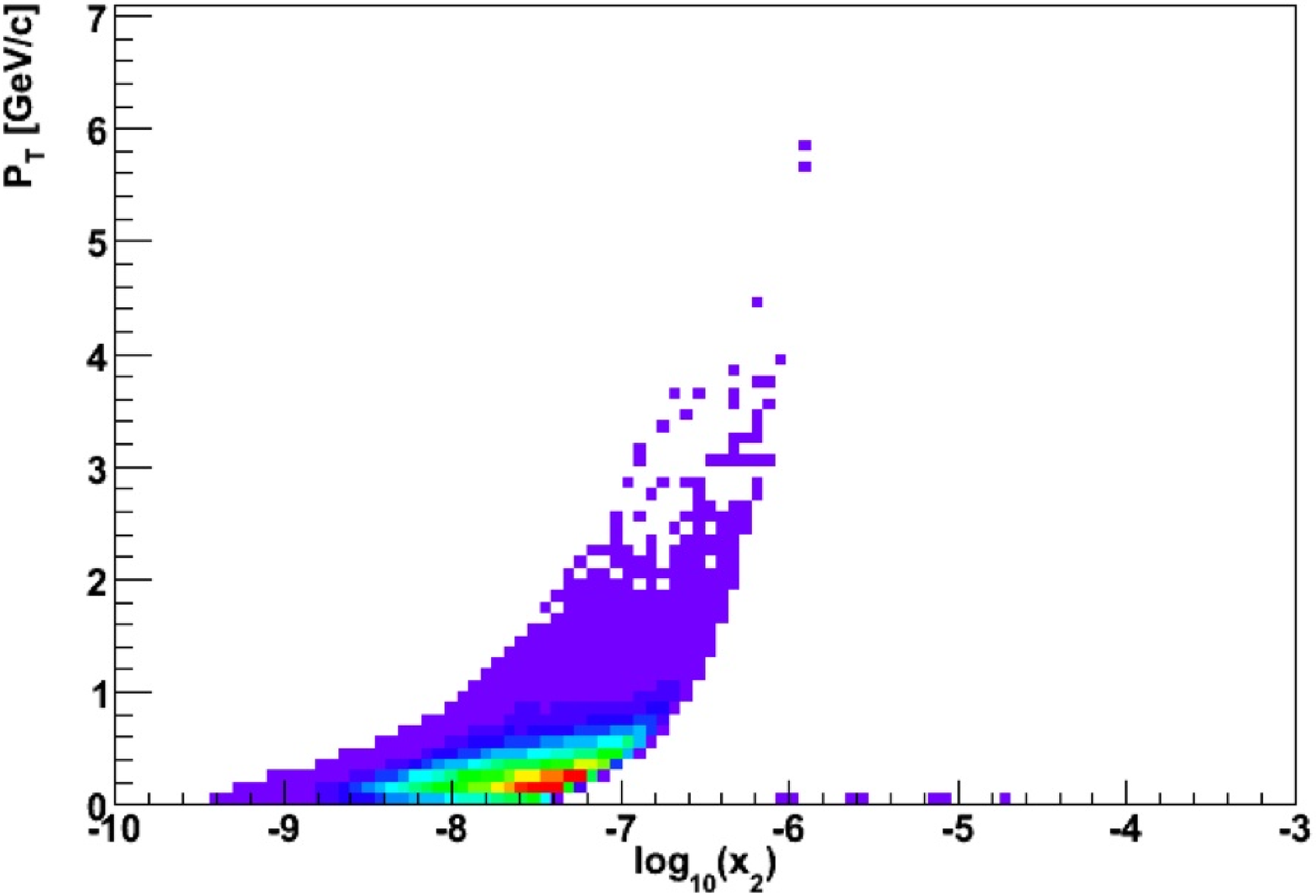}
\caption{
\label{fig:lowx}
(left) Reconstruction of $\pi^0$ and $\eta$ particles via 2-photon
decay into the ATLAS ZDC.
(right) Acceptance of ATLAS ZDC for $\pi^0$'s in $p_T$ vs. $x_2$.
}
\end{center}
\end{figure}

The ATLAS ZDCs will be primarily used for centrality selection
in A+A as well as the study of ultraperipheral collisions (which
will explore similar physics as next-generation electron ion
colliders)~\cite{Strikman:2005yv}.  
However, their ability to reconstruct far-forward
$\pi^0$'s in $p+p$ collisions~\cite{ATLASZDC}, 
shown in Fig.~\ref{fig:lowx}, 
gives them particular utility in addressing low $x$ physics.
Via the kinematic relations typical for CGC physics, 
$x_2 \sim (p_T/\sqrt{s})e^{-y}$ (where $p_T$ and $y$ are 
the transverse momentum and rapidity
of the detected $\pi^0$), 
Fig.\ref{fig:lowx} shows the effective
range in $x_2$ and $p_T$ reached for $\pi^0$'s reconstructed in
the ZDC.  By probing $x_2$ down to $10^{-6}-10^{-8}$ at 
moderate $p_T$, even $p+p$ collisions will become interesting
laboratories to study universal features of hadronic wave 
functions~\cite{Iancu:2003xm}.

\section{Conclusion}
In conclusion, ATLAS is preparing intensely for heavy ion 
data at the LHC in 2008 and beyond.  Studies shown in this
work include bulk observables in p+p and A+A, 
inclusive jets in p+p and A+A, quarkonia reconstruction,
and early steps towards low-$x$ physics and ultraperipheral
collisions.  Of course, important work remains to be done
on many tasks.  New collaborators
are always welcome, to work on software, analysis, physics and
trigger issues!

\ack
The author would like to thank the Quark Matter 2006 organizers
and ATLAS management for the invitation to speak in Shanghai.
Thanks as well to colleagues in the ATLAS Heavy Ion Working Group for
providing the physics and detector studies, as well as 
invaluable advice and comments on this manuscript.
This work was supported in part
by the Office of Nuclear Physics of the U.S. Department of Energy under
contracts: DE-AC02-98CH10886.

\end{document}